**Title:**
Neural and phenotypic representation under the free-energy principle

**Authors:**
Maxwell J. D. Ramstead[1,2,3] (maxwell.ramstead@mail.mcgill.ca)
Casper Hesp[3,4,5,6] (c.hesp@uva.nl)
Alexander Tschantz[9,10] (tschantz.alec@gmail.com)
Ryan Smith[7] (rsmith@laureateinstitute.org)
Axel Constant[2,3,8] (axel.constant.pruvost@gmail.com)
Karl Friston[3] (k.friston@ucl.ac.uk)

**Affiliations:**
1. Division of Social and Transcultural Psychiatry, Department of Psychiatry, McGill University, Montreal, Quebec, Canada.
2. Culture, Mind, and Brain Program, McGill University, Montreal, Quebec, Canada.
3. Wellcome Centre for Human Neuroimaging, University College London, London, UK, WC1N3BG.
4. Department of Psychology, University of Amsterdam, Science Park 904, 1098 XH Amsterdam, Netherlands.
5. Amsterdam Brain and Cognition Centre, University of Amsterdam, Science Park 904, 1098 XH Amsterdam, Netherlands.
6. Institute for Advanced Study, University of Amsterdam, Oude Turfmarkt 147, 1012 GC Amsterdam, Netherlands.
7. Laureate Institute for Brain Research, Tulsa, Oklahoma, USA.
8. Theory and Method in Biosciences, Level 6, Charles Perkins Centre D17, Johns Hopkins Drive, University of Sydney, NSW 2006 Australia.
9. Sackler Centre for Consciousness Science, University of Sussex, Brighton, UK
10. Department of Informatics, University of Sussex, Brighton, UK

**Abstract:**
The aim of this paper is to leverage the free-energy principle and its corollary process theory, active inference, to develop a generic, generalizable model of the representational capacities of living creatures; that is, a theory of phenotypic representation. Given their ubiquity, we are concerned with distributed forms of representation (e.g., population codes), whereby patterns of ensemble activity in living tissue come to represent the causes of sensory input or data. The active inference framework rests on the Markov blanket formalism, which allows us to partition systems of interest, such as biological systems, into internal states, external states, and the blanket (active and sensory) states that render internal and external states conditionally independent of each other. In this framework, the representational capacity of living creatures emerges as a consequence of their Markovian structure and nonequilibrium dynamics, which together entail a dual-aspect information geometry. This entails a modest representational capacity: internal states have an intrinsic information geometry that describes their trajectory over time in state space, as well as an extrinsic information geometry that allows internal states to encode (the parameters of) probabilistic beliefs about (fictive) external states. Building on this, we describe here how, in an automatic and emergent manner, information about stimuli can come to be encoded by groups of neurons bound by a Markov blanket; what is known as the *neuronal packet hypothesis*. As a concrete demonstration of this type of emergent representation, we present numerical simulations showing that self-organizing ensembles of active inference agents sharing the right kind of probabilistic generative model are able to encode recoverable information about a stimulus array.

**Acknowledgements:**
We are grateful to Thomas Parr, Alex Kiefer, Inês Hipólito, Laurence Kirmayer, Conor Heins, and Yan Yufik for helpful comments and discussions that contributed to our work on this paper. Researchers on this article were supported by a Social Sciences and Humanities Research Council postdoctoral fellowship (MR), a NWO Research Talent Grant from the Dutch Government (Ref: 406.18.535) (CH), by a PhD studentship from the Sackler Foundation and the School of Engineering and Informatics at the University of Sussex (AT), by the William K. Warren Foundation (RS), by the Australian Laureate Fellowship project A Philosophy of Medicine for the 21st Century (Ref:  FL170100160), by a Social Sciences and Humanities Research Council doctoral fellowship (Ref: 752-2019-0065) (AC), and by a Wellcome Trust Principal Research Fellowship (Ref: 088130/Z/09/Z) (KJF). AT is grateful to the Dr. Mortimer and Theresa Sackler Foundation, which supports the Sackler Centre for Consciousness Science.

## 1. Introduction

Living creatures are endowed with a representational capacity: they behave as if states of their bodies encode information or probabilistic beliefs about salient features of their environment, which they seem to leverage to generate contextually appropriate, adaptive behaviour (Egan, 2018; Kiefer & Hohwy, 2018; Ramstead, Friston, & Hipólito, 2020; Ramstead, Kirchhoff, & Friston, 2019). Most multicellular creatures have evolved specialised cell populations – nervous systems, comprising neural and nonneural (e.g., glial) cells – that are dedicated to realizing this representational capacity. Through continuous tuning of network connections, groups of neurons self-organize to form functionally integrated ensembles that look as if they are recapitulating the statistical structure of the environment, in a manner that organisms can leverage to generate adaptive patterns of behaviour.

Population codes, where neural ensembles look as if they encode probabilistic beliefs about the statistical structure of the environment (i.e., probability densities over the latent causes of sensory data), are as ubiquitous as they are well documented (Pouget, Dayan, & Zemel, 2000). Single neurons are known to respond preferentially to specific kinds of stimuli; e.g., neurons in primary visual cortex respond preferentially to specific orientations and luminance contrasts of a stimulus presented in their receptive field. Neurons also look as if they work together to infer the most likely causes of sensory input in distal sensory epithelia, such that stimulus information can only be fully recovered by examining activation patterns across many cells. This has been shown extensively in the visual system (Usrey & Reid, 1999). It is well known, for instance, that neurons in primary visual cortex look as if they infer the causes of stimuli (i.e., the patterns of photons) that impinge upon retinal neurons (Maunsell & Van Essen, 1983); that is, the average activity pooled over these neural populations looks as if they track features of the stimulus presented to retinal neurons. Something similar can be said of many other neural populations, e.g., place cells in the hippocampus (Bair, 1999; Borst & Theunissen, 1999) and topographically organized neurons in motor cortex (Tolhurst, Movshon, & Dean, 1983).

How are these coordinated feats of representation possible? How does specialised neural tissue, and indeed any kind of phenotypic state, come to encode information about the world in which the organism is embedded? Is it possible to formulate a general theory of how the confluence of brains, bodies, and their behaviours can represent the environment? Can we construct a general theory of phenotypic representation?

In this paper, we leverage the apparatus of the free-energy principle and its corollary process theory, known as active inference, to propose a general theory of neural representation. This framework opens, more broadly, onto a theory of phenotypic representation, where the brains, bodies, and behaviours of organisms – and indeed, the coordinated behaviour of groups of organisms – are cast as engaging in inference about the causes of their sensory states at several nested spatial and temporal scales.

The free-energy principle is a novel multiscale framework for studying belief-driven adaptive action (Friston, 2020). The free-energy principle provides us with a theory derived from first principles regarding what it means to exist as a system with a phenotype (i.e., to exist in a nonequilibrium steady state regime). On this account, to exist as a living system entails continually revisiting the neighbourhood of the same set of states (e.g., remaining within a certain range of body temperatures or ecological environments). Technically, such systems are endowed with a random dynamical or pullback attractor. This engenders a nonequilibrium steady state density that we can associate with the phenotype of a living system. Active inference, in turn, provides us with a (Bayesian) mechanics that explains how organisms remain within their phenotypic bounds (Ramstead, Badcock, & Friston, 2018). The framework rests on the Markov blanket formalism, which allows us to partition systems of interest into internal states, external states, and the blanket (active and sensory) states that render internal states conditionally independent of external states. This fundamental nonequilibrium Markovian structure (Friston, Wiese, & Hobson, 2020) will guide our investigations here. We argue that representational capacities of living creatures can be formalized in a biologically realistic way via the active inference framework. We develop a generic and generalizable model of the representational capacities of living creatures or, indeed, particles with *particular* states (namely, internal states and their Markov blanket).

We draw on one prescient operationalization of this distributed representational capacity – the construct of a *neuronal packet* – and provide a proof of principle for the *neuronal packet hypothesis* (Yufik, 1998; Yufik & Friston, 2016). According to the neuronal packet hypothesis, cortical neurons perform inference about the causes of patterns in sensory neuron stimulation by forming neuronal packets, which are neural ensembles wrapped in a Markov blanket, such that neuronal packets come to represent recoverable features of the stimulus. In one sense, the notion of a neuronal packet can be read as homologous to the notion of a neuronal group, ensemble or assembly that has a long pedigree in neurology and neuroscience (Buzsaki, 2010; Edelman, 1993; Edelman, 1998; Gerstein & Kirkland, 2001; Hebb, 1949; Mountcastle, 1997; Nicolelis & Lebedev, 2009; Sherrington, 1911; von der Malsburg, 1981).

We argue that the representational capacity of creatures is a consequence of their Markovian structure. The Markovian partition of a system in a nonequilibrium steady state regime equips the internal states of a system with a dual-aspect information geometry. Internal states have an intrinsic information geometry that describes their trajectory over time in state space, as well as an extrinsic geometry based on probabilistic beliefs about fictive external states (that is, beliefs that may or may not correspond to true external states), encoded by internal states. Here, we show simulation results that concretely demonstrate in silico that – when suitably configured under the right probabilistic model – the self-organizing ensembles of active inference agents (ensembles of simulated neurons) that share the same model are able to encode recoverable information about a stimulus array. We argue that the configurations at which the system arrives through such self-organised coordination can be cast as a form of embodied inference (i.e., arriving at posterior state estimates over the causes of sensory stimulation).

Although we focus on neuronal ensembles, it is important to highlight that the account that we develop here, based as it is on the free-energy principle, can be applied to any group of living (sub-)systems that entertain a shared generative model. This means that the representational capacity that we discuss in this paper is general. It need not be restricted to cells, or even to animals, and has the resources to account for potential collective representational capacities across all kinds of self-organising systems; e.g., plant dynamics (Calvo & Friston, 2017).

Extant work on this topic has focused on the representational capacities of individual systems (Friston, 2013; Kuchling, Friston, Georgiev, & Levin, 2019), sometimes composed of other systems (Palacios, Razi, Parr, Kirchhoff, & Friston, 2020). In this paper, we extend this line of reasoning to *patterns of group activity* that enable a group of agents (be they neurons or other elements of larger systems that realise the appropriate dynamics) to *collectively represent* features of an external environment. This extension is crucial for understanding the intentional capacities of living systems, where hierarchical and distributed dynamics are ubiquitous.

## 2. Background: Multiscale active inference

### 2.1. Active inference and Markov blankets

The free-energy principle begins from the observation that living creatures tend to revisit the neighbourhood of the same states of being (Ramstead et al., 2018). This is formalized as the claim that the dynamics of living creatures are underwritten by a nonequilibrium steady state (Friston, 2020). Living systems exist far from thermodynamic equilibrium, in a regime of characteristic states that we associate with a nonequilibrium steady-state density over the system's states, with high probability states corresponding to typical states given 'the kind of creature that I am', i.e., to those that are consistent with an organism's phenotype. The free-energy principle explains how living creatures remain alive (i.e., away from equilibrium), by continuously generating patterns of adaptive action that bring about preferred sensory data or observations compatible with their survival (Friston, 2020). Thus, according to active inference, to exist as a living thing means to continually produce sensory evidence of one's own existence (i.e., evidence that 'I am in or near a nonequilibrium steady state); what is sometimes called self-evidencing (Hohwy, 2016).

Active inference rests on a few fundamental, interlocking mathematical constructs: these comprise the nonequilibrium steady state densities just discussed, as well Markov blankets, generative models, and variational free-energy (Ramstead, Constant, Badcock, & Friston, 2019; Ramstead et al., 2020; Ramstead, Kirchhoff, et al., 2019). A growing literature appeals to Markov blankets to account for biological self-organization and the ability of living systems to remain alive over time (Constant, Ramstead, Veissière, Campbell, & Friston, 2018; Friston, 2013, 2020; Hipólito, 2019; Kirchhoff, Parr, Palacios, Friston, & Kiverstein, 2018; Kuchling, Friston, Georgiev, & Levin, 2019; Palacios, Razi, Parr, Kirchhoff, & Friston, 2020; Ramstead et al., 2018; Ramstead, Constant, et al., 2019). This scheme gives us a principled definition of what it means to exist as a system (Kirchhoff et al., 2018; Ramstead et al., 2018). Minimally, to exist as a system implies a form of conditional independence with respect to the rest of the world; and Markov blankets operationalize this independence.

A Markov blanket defines a system of interest by identifying the states that mediate its probabilistic relations with the events and regularities of the environment in which it is embedded, and which cause or generate those sensory states. By inducing conditional independence (Pearl, 1988), Markov blankets enable us to define the boundaries between a system and its environment, and thereby delimit the system as such (Friston, 2013, 2020; Friston, Levin, Sengupta, & Pezzulo, 2015). We define a system of interest, delimiting it from the system in which it is embedded (e.g., a living creature in its ecological niche, or a cell in the intercellular milieu), by defining a third set of states – the Markov blanket itself – that mediate the causal influences between them. The Markov blanket is further partitioned into sensory and active states: active states affect, but are not affected by, external states; whereas sensory states affect, but are not affected by, internal states. Given this partition, internal states are independent of external states, given blanket states. See Figure 1.

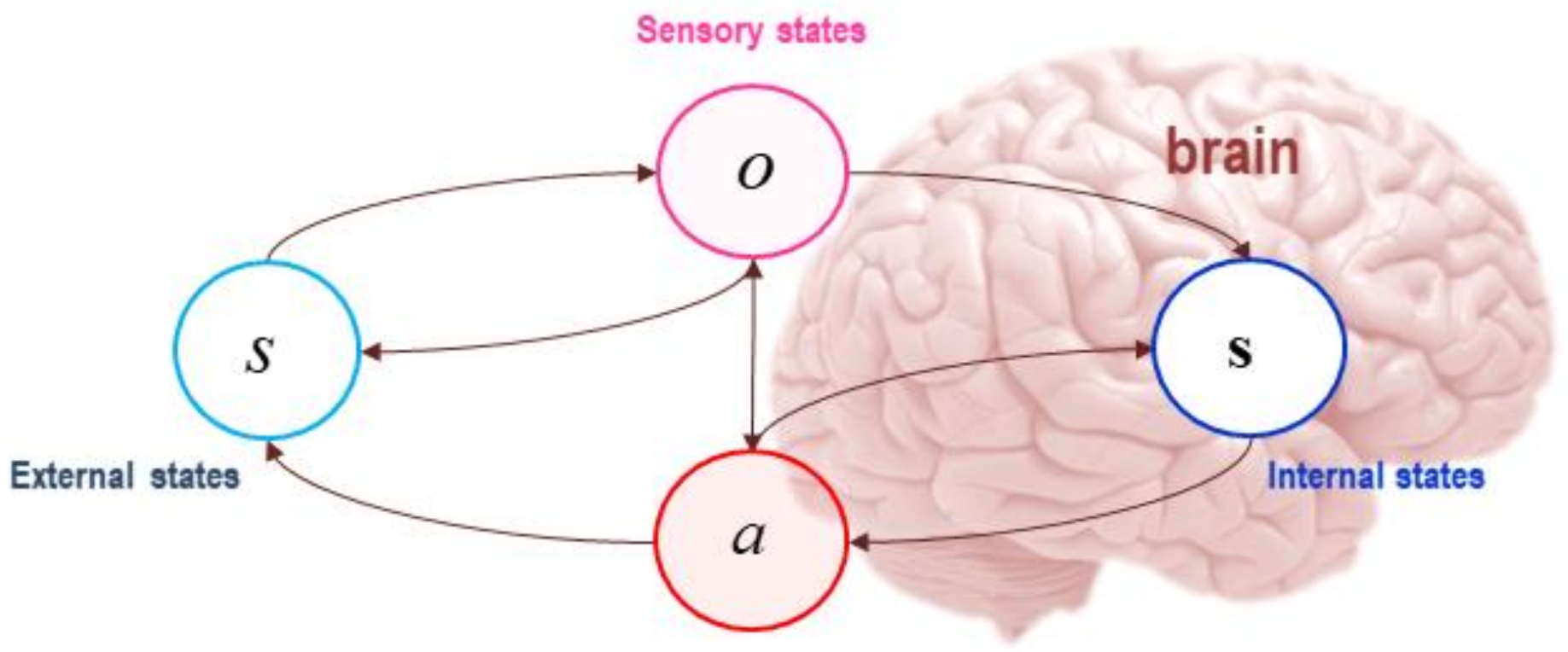

**Figure 1.** Markov blankets under active inference

This figure depicts the statistical relations that define the Markov blanket. The existence of a Markov blanket induces certain conditional independences: the presence of the blanket partitions the system into internal states (denoted **s**), external states (denoted *s*), and the blanket states themselves – which render internal and external states independent of each other, when conditioned on blanket states. The Markov blanket itself comprises active states (*a*) and sensory states (*o*). From Ramstead, Kirchhoff, et al. (2019).

## 2.2. Generative models, variational free-energy, and the mechanics of inference

Active inference is a formulation of *Bayesian mechanics*, which tells us about the flow of internal and active states and how this keeps the Markov blanket in play (i.e., keeps the boundaries that separate an organism from its environment intact). For action and perception to be adaptive and knowledge-driven, some structures must exist that harness this knowledge in a form amenable for use in guiding adaptive action. These structures are the internal states of creatures, shielded behind their Markov blankets. Active inference allows us to cast perception and action (i.e., the flow of internal and active states, respectively, in response to perturbations from sensory states) as maximising the evidence for an implicit statistical model, which living creatures embody and which guides their production of adaptive action through the selection of (beliefs about) sequences of actions, or policies (Friston, 2020).

Formally, generative model is a joint probability density over the variables of a system; in the case of active inference, this a joint probability density over organismic states (so-called particular states, i.e., internal, sensory, and active states) and fictive external states (Ramstead, Kirchhoff, et al., 2019). Under the free-energy principle, the generative model is just an interpretation of the nonequilibrium steady state density that we associate with the organism’s phenotype (Friston, 2020). Heuristically, we can either look at this density from the point of view of the physics of flow, as the underlying manifold that structures the flow of states; or as a joint probability density over all systemic states (Ramstead et al., 2020). In brief, the presence of a Markov blanket in a system at nonequilibrium steady state means that we can look at the system's internal states as if they parameterize beliefs about (probability distributions over) external states. Because the external states now play the role of random variables they are often referred to as the latent or hidden states (i.e., hidden behind the Markov blanket). These hidden states can be read as explanatory fictions that the organism entertains to explain how its data were caused (Ramstead et al., 2020).

A modest representational capacity follows from this Markovian, nonequilibrium setup (Friston, 2020; Friston, Wiese, et al., 2020; Parr, Da Costa, & Friston, 2020). Indeed, the Markovian structure just described leads to a dual-aspect information geometry that licences a description of internal states where they encode a conditional (Bayesian) belief about the external world. Information geometry is essentially concerned with measurement of spaces that represent (the sufficient statistics of) probability distributions. Simply put, the existence of the Markovian partition implies that, for any blanket states of a system, one can identify (an average over) internal states and a corresponding conditional belief (i.e., probability density) about external states. Thus, in addition to the 'intrinsic' information geometry of the system – which captures the probability of internal states over time – the Markovian structure endows the system with an 'extrinsic' information geometry, which captures beliefs about concurrent external states (Friston, Wiese, et al., 2020).

According to the free-energy principle, living systems enact dynamics (patterns of action and perception) that entail a generative model (Friston, 2012; Ramstead, Kirchhoff, et al., 2019). What this means is that, mathematically, the behaviour looks as if it was produced by a specific causal process (a specific generative process), the structure of which is recapitulated in the generative model. In producing adaptive behaviour, living systems will look as if they are performing inference about the causal structure of their ecological niche, and maximising the evidence for a model of the world that they implicitly embody. More specifically, they do so by optimizing a bound on the (negative log) of this evidence known as the variational free-energy – which the organism can compute from the dynamics of its sensory states. By minimizing variational free-energy (a.k.a. maximizing model evidence), internal states of an organism come to encode a probabilistic 'best guess' about the causes of sensory input (Kiefer & Hohwy, 2018; Ramstead, Kirchhoff, et al., 2019). This is known as the (variational) free-energy principle (Friston, 2010, 2020).

To summarize, the entailment of a generative model by adaptive behaviour, and the concomitant partition of the system into internal and external states separated by blanket states (and thereby the maintenance of the Markov blanket), leads to the induction of beliefs or knowledge about external states, parameterized by internal states. The physical states of the system then come to encode (the sufficient statistics) of distinct posterior beliefs about latent states. These are explanatory fictions or hypotheses about what is causing the sensory states. This means that the physical quantities that organisms embody encode the (parameters of) posterior or conditional density (i.e., Bayesian beliefs) over hidden or latent states that might best explain the manner in which sensory data were generated (Ramstead, Kirchhoff, et al., 2019). Theexistence of a partition into internal, blanket, and external states thus also induces beliefs or knowledge about external states encoded by internal states and leads to the emergence of representational capacity in living creatures (Friston, 2012; Friston, 2020; Friston, Wiese, et al., 2020; Ramstead et al., 2020). The ensuing Bayesian mechanics suggests a deep if somewhat deflationary connection between existence and knowledge; namely, to exist is to know – in the sense of holding Bayesian beliefs (John Campbell, personal communication).

### 2.3. Nested Markov blankets and multiscale self-organization

This Markovian structure of living systems is iterated recursively across all levels of biological organization (Friston, 2020; Kirchhoff et al., 2018; Palacios et al., 2020). Since most living systems are composed of parts that are themselves living systems (e.g., organisms are made of organs, which are made of cells), the Markovian structure iterates recursively, allowing us to explain the dynamics of any system that exists at nonequilibrium steady state across levels of description – from cells to societies (Friston, 2020; Kirchhoff et al., 2018; Ramstead et al., 2018; Ramstead, Constant, et al., 2019). It is thus blankets of blankets, as it were, all the way up, and all the way down. See Figure 2.

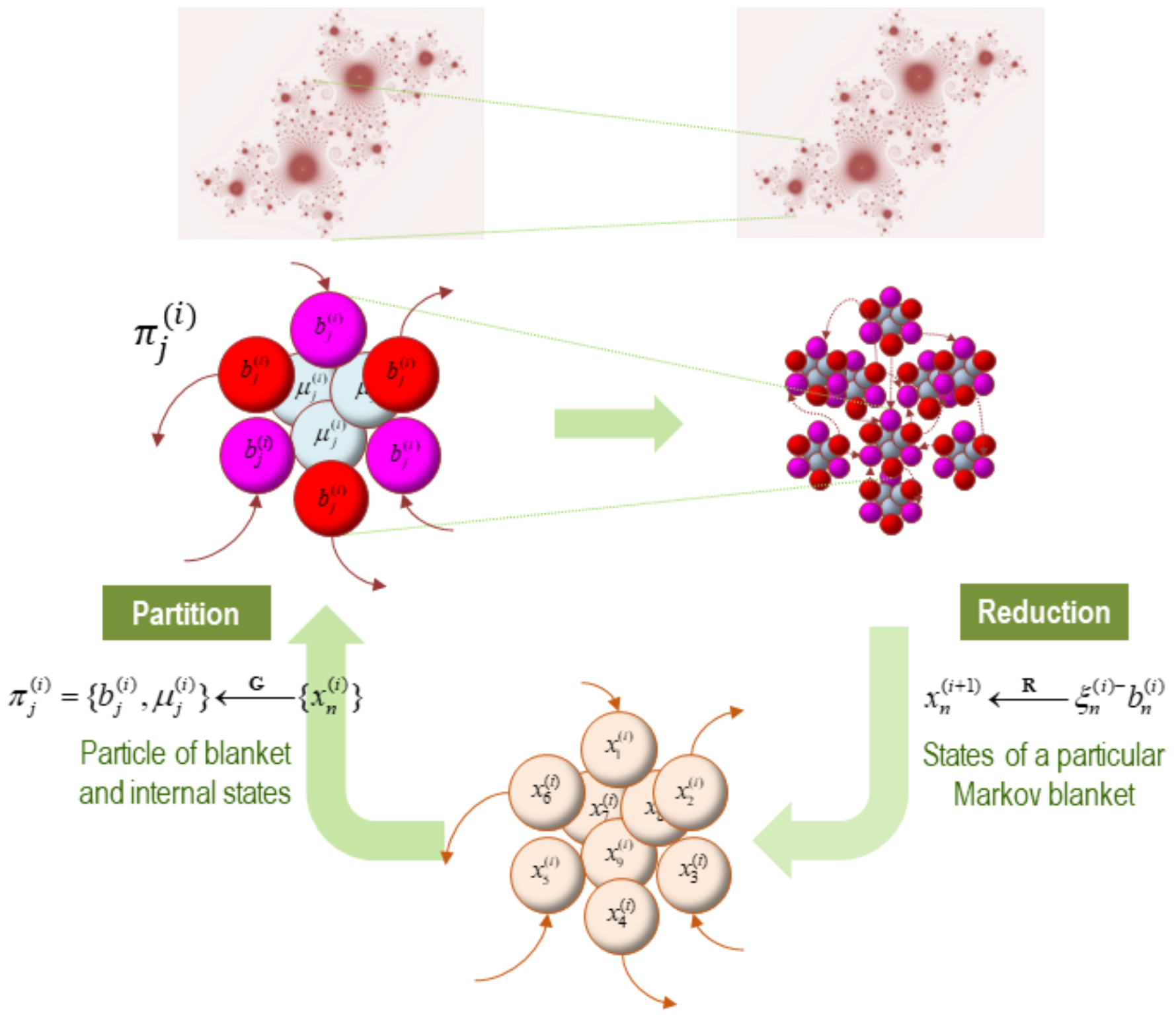

**Figure 2.** Nested Markov blankets

This figure depicts the recursive structure of nested Markov blankets. At each successive level of the nested hierarchy, superordinate dynamics at successively larger and slower scales emerge from, and constrain, dynamics arising at subordinate, smaller and faster scales. We can read the figure starting with the bottom panel. This panel depicts an ensemble of nine states – denoted $x$ – which have not yet been differentiated into the partition prescribed by a Markov blanket. The formalism deploys renormalization group theory to derive a scale-free dynamics (Friston, 2020). In brief, we start at the bottom panel with a group of vector states (here, nine are depicted). We use a grouping operator **G** to group these states according to their conditional independences, by computing the Jacobian of their adjacency matrix. This leaves us with a particular partition of the same states, now grouped into blanket (sensory and active) and internal states. We then use a dimension reduction operator **R** to derive dynamics at the next scale. Here, the fast and stable modes of the blanket states and the internal states are dropped. The principal eigenstates of each blanket are taken to define new vector states at the scale above; and the process repeats until all states have been accounted for. Depicted in the upper panels are particles, which are defined in terms of the conditional dependences between these states and that define the Markov blanket of each particle. Here, a particle $\pi$ is a mixture of blanket states ($b$) and internal states ($\mu$). From Friston (2020).

The strategy deployed by proponents of active inference to model self-organization at several scales is to show that some higher-level phenomenon (e.g., pattern formation) emerges from the coordinated inference dynamics of component phenomena at the subordinate scale, given that they share a generative model (Friston, 2013; Kirchhoff et al., 2018). In this work, each individual agent is equipped the same generative model (which is a metaphor for a shared genetic code). The parameters of this generative model encode knowledge about the kinds of signals that different cell types receive and send, as a function of the kind of cell they are. (We discuss this in detail in the Methods and Results sections below). Based on this shared knowledge, each cell in the ensemble is able to infer its place and its role (as an internal, active, or sensory state) in the higher-order system, as a function of the evidence that each cell accumulates that is indicative of such a role, in terms of probabilistic relations between roles as defined by the Markovian partition – e.g., internal states do not communicate directly with other internal states (which are shrouded behind their blankets), but can communicate with blanket states, etc.

In other words, an ensemble self-organizes to some target configuration because its components share the same expectations about the typical observations or data that are generated by the behaviour of it and other members of the ensemble. For a system of units sharing a generative model, the free-energy minimum for each cell coincides with the free-energy minimum for the ensemble (Friston, 2013; Friston et al., 2015; Kuchling et al., 2019; Palacios et al., 2020). Target morphologies emerge as the result of group inference operating on a shared generative model; where each unit comes to ‘know its place’ by inferring what role it must be playing, given the signals it is receiving from other units and those it sends.

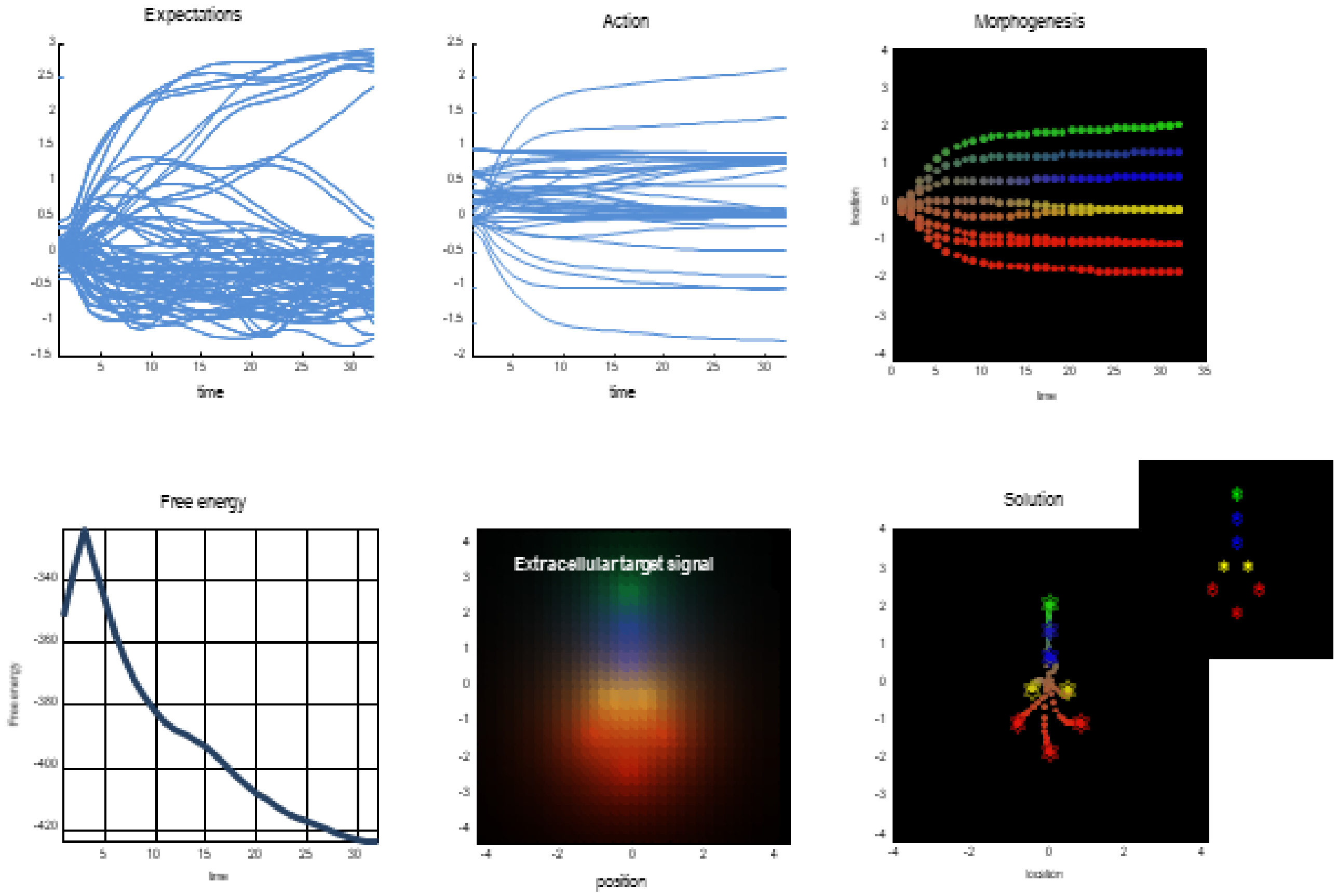

**Figure 3.** Morphogenesis under the free-energy principle.
This figure depicts the numerical results reported in Friston et al. (2015), which were expanded upon by Palacios et al. (2020) and by Kuchling et al. (2019), and which are described further in the text. The bottom right panel depicts the target configuration and the movement of cells towards this configuration, and the central bottom panel depicts the signalling profile associated with each target position, which each cell uses to find its place. The top panels display the expectations of each cell as it moves towards the final position. The bottom left panel depicts the free-energy of the ensemble. Free-energy increases initially as each cell figures out what kind of cell it is, and then drops to the global minimum, which coincides with each cell finding its place in the target configuration.

## 2.4. Markov blankets in the brain and the neuronal packets hypothesis

As discussed above, Markov blankets have been used to define the boundaries and dynamics of systems interacting across scales (Friston, 2020; Hipólito, 2019; Kirchhoff et al., 2018; Palacios et al., 2020; Ramstead et al., 2018; Ramstead, Constant, et al., 2019). This applies to the brain as well (Friston, Fagerholm, et al., 2020; Hipólito et al., 2020; Palacios, Isomura, Parr, & Friston, 2019; Parr & Friston, 2018). To start with, single neurons have their Markov blankets (Kiebel & Friston, 2011). The sensory states of the neuron are the synapses through which it receives signals from other cells. The active states of neurons are their axons and synaptic terminals through which they convey action potentials and patterns of neurotransmitter release onto other cells. Each neuron has a preferred receptive field profile or response vector that represents its preferred stimulus type, which we can associate with the afferent synaptic connections of the neurons that encode its likelihood function (i.e., that encode what inputs are those that provide evidence for the hypothesis it represents). Active inference can be used to explain neural behaviour at three different scales: the fast timescale of inference (changes in activation or depolarization), the slower timescale of associate learning (changes in synaptic efficacy or gain), and the even slower timescale of structure learning (structural change in dendritic architecture) (Kiebel & Friston, 2011). Empirically, there is emerging evidence for the self-evidencing behaviour of neuronal ensembles in terms of dendritic architectures; even in cell cultures (Isomura & Friston, 2018).

On the active inference account, the architecture of (nested) Markov blankets discussed above might form the basis of brain organization generally (Friston, Fagerholm, et al., 2020; Hipólito et al., 2020). The causal factors that figure in generative models are thought to correspond to anatomically segregated brain areas (Friston & Buzsáki, 2016; Parr & Friston, 2018). The evidence once thought to suggest modular processing in the brain may be better interpreted as evidence for structures enabling this factorization. In other words, functionally specialised brain regions may encode factors of a factorized probability density, which are conditionally independent but related to each other through their mean fields (Friston, Parr, & de Vries, 2017; Parr, Sajid, & Friston, 2020). This means that neuronal populations may encode distinct factors, each segregated by their Markov blanket (e.g., the identity of an object vs. its spatial location).

This leaves open the question of how *neural ensembles* come to encode information. The *neuronal packet hypothesis* (Yufik & Friston, 2016) proposes that neurons are able to encode information about external causes by forming coherent, bounded ensembles, enshrouded behind a Markov blanket. The self-organization of a Markovian ensemble of neurons in response to perturbations (action potentials) coming from sensory neurons can thus be interpreted as a form of posterior inference, with each packet encoding a different hypothesis (Yufik, 2019; Yufik & Friston, 2016). This is broadly consistent with theories of structural representation, according to which neurons encode exploitable information about a target domain by mirroring the (mathematical or statistical) structure of that target domain (Kiefer & Hohwy, 2018, 2019; Kiefer, 2020). The neuronal packet hypothesis goes further and proposes that the capacity of an organism to infer and interpret the causes of its sensory states results from the binding together of neuronal populations through the formation of a boundary (i.e., a Markov blanket) around the neuronal packet. To 'mobilise' such a packet – in response to sensory perturbation – would be to understand a world composed of interacting causes (Yufik, 2019; Yufik & Friston, 2016).

Empirical evidence supports the neuronal packet hypothesis, suggesting that neurons process information by forming task-dependent, transient coalitions on the fly; for an extensive review of the empirical evidence for this type of 'neural reuse' (Anderson, 2014). Thus, brain activity arises from coalitions of neurons functioning together as a cohesive unit, in a context- and goal-dependent manner. This allows for the combination of information about more or less independent causal factors encoded by the physical states of brain networks (Parr, Sajid, et al., 2020). Neurons are thus trying to form functional coalitions dynamically, to bring to bear the factors that they encode in a multifactorial coordinated inference; see also Park and Friston (2013). Strikingly, evidence for the Markovian structure of brain dynamics comes from recent work in neuroimaging, which demonstrates that brain regions form Markovian 'parcels' (Friston, Fagerholm, et al., 2020).

In recent active inference models, neurons are endowed with the prior belief (encoded in their shared generative model) that, in the absence of external perturbations, their signalling profiles will align or synchronise (Palacios et al., 2019). This might provide the initial impetus for packet formation. Each neuron is equipped with a generative model that allows it to infer its functional identity (or role in the Markovian partition) in the superordinate ensemble. It thus follows that perceptual inference arises as the neuronal system self-organizes into coherent packets in response to sensory perturbations.

One further phenomenon of interest that may be worth considering in this context is structural self-organization – and associated functional segregation (Zeki, 2005) – during early development and neuronal differentiation. Instead of neurons being genetically hard-wired to come to represent specific stimuli, some work suggests that common patterns of functional segregation may be more weakly specified by the inherited constraints on large-scale axonal structure (e.g., the path of optic nerve signals to visual cortices via the thalamus), where the functional role of a neuron (or of a locally connected population of neurons) instead emerges from self-organization in response to the statistics of the signals it receives early in development (Bienenstock, Cooper, & Munro, 1982; Peters & Yilmaz, 1993; Sin, Haas, Ruthazer, & Cline, 2002; Sur, Garraghty, & Roe, 1988; Sur & Rubenstein, 2005; von der Malsburg, Phillips, & Singer, 2010; Von Melchner, Pallas, & Sur, 2000; Zeki & Shipp, 1989). This has been most starkly demonstrated in studies of experimentally manipulated large-scale neuronal pathways, e.g., studies that redirect visual signals from the optic nerve to the medial geniculate nucleus and auditory cortices. Remarkably, in these cases, populations of auditory cortex neurons become sensitive to specific visual properties similar to those normally seen in visual cortex. Other studies of developmental windows, such as work showing selective blindness to particular visual features (e.g., vertical contours) not present in the early environment, and associated pruning of the necessary synaptic connections to implement the inference of such properties, further illustrates this point (Blakemore & Cooper, 1970; Buonomano & Merzenich, 1998; Huang & Reichardt, 2001; Muir & Mitchell, 1973; Rice & Barone, 2000; Singer, Freeman, & Rauschecker, 1981; Thoenen, 1995). Thus, there is a strong sense in which (under broad genetic constraints on large-scale axonal wiring patterns) neural populations infer what specific functional role they play based on the early inputs they receive.

## 2.5. Neural representation in the active inference framework

This paper builds on the technical developments of active inference models of hierarchical self-assembly and self-organization (Friston et al., 2015; Kirchhoff et al., 2018; Palacios et al., 2019; Palacios et al., 2020). This work has shown that a Markov blanket emerges naturally around a group of active inference agents that share the right kind of generative model (Friston et al., 2015; Kirchhoff et al., 2018; Kuchling et al., 2019). The strategy deployed in these papers is to recast the problem of superordinate boundary formation – the emergence of a Markov blanket – as the consequence of inference conditioned on a shared generative model.

Extending this work, our strategy in this paper is to explain the self-organization of neural tissue into neuronal packets at a superordinate level by modelling coordinated inferences of single neurons sharing a generative model. The technical novelty in our setup is that each individual neuron must now infer to which higher-order Markov blanketed system it belongs, in addition to the role that it plays in that superordinate system. This role in the superordinate ensemble is implemented by leveraging the partition of the system into different kinds of states, which follows from the existence of a Markov blanket at the superordinate scale (and beliefs about membership in that blanketed system): individual neurons must infer whether they play the role of a sensory state, an active state, or an internal state, where each of these roles is defined in terms of the conditional independencies that define the Markov blanket.

We model neuronal packet formation as the result of inferences about membership to a higher-order Markov blanketed ensemble, which can be cast as a form of inference. Each neuron has an individual response vector; that is, each neuron responds preferentially to some features being present in their receptive field. When sensory neurons detect their preferred stimulus in their receptive field, they send signals (i.e., action potentials) to cortical neurons, e.g., neurons in the primary visual context receiving signals from sensory neurons in the retina (through the lateral geniculate nucleus). We model the later type of neuron in this work. We hypothesize that cortical neurons use these signals as evidence for a specific inference: *that they are the internal states of a neuronal packet at the superordinate level*. Neurons exchange signals as well, which express their beliefs about blanket membership at the superordinate scale. This group inference and communication leads to the formation of a neuronal packet, which encodes a hypothesis about what might have caused the sensory impressions on visual receptors. Neuronal packet membership can be seen as a measure of evidence for that hypothesis. Thus, neuronal packet formation corresponds to the self-organization of neural ensembles enabled by inferences about membership to a higher-order Markov blanketed system. This provides a generalizable mechanism for distributed neural representations, premised on a nonequilibrium, Markovian, and recursively nested architecture.

In short, we hypothesize a specific architecture that links neurons in the sensory epithelia to neurons in the brain. The response vector of sensory neurons encodes the kinds of target features in their receptive fields to which they are most sensitive (e.g., orientation and contrast). This sensitivity means that, in the presence of their preferred stimulus, the rate at which action potentials are generated increases. Our hypothesis is that, when they detect increases in particular signalling patterns from sensory neurons, cortical neurons take this as evidence to *infer that they are internal states of a higher-order ensemble*. Essentially, on this account, the action potentials generated by each cortical neuron broadcast to other cortical neurons a signal that conveys its beliefs about whether it is an internal state of a superordinate ensemble.

## 3. Methods

To concretely demonstrate our proposed formulation, we simulate $N = 256$ 'particles' operating in a two-dimensional phase space, where the location of a particle is denoted $x \in \mathbb{R}^2$ and the equations of motion are largely based on those used by Palacios and colleagues (2020). Each particle – representing an individual neuron – is modelled as an active Bayesian agent, partitioned into sensory $s$, active $a$ and internal $\mu$ states. The internal states parameterize beliefs about (i.e., a probability density over) hidden external causes of sensory input and are optimised using Gaussian filtering (Palacios et al., 2020). In other words, they are updated in order to minimise variational free-energy. In the current context, the internal states parameterize beliefs about the identity of the particle $\psi \in \mathbb{R}^4$, where each identity corresponds to membership in a superordinate ensemble.

Particles emit signals $\psi \in \mathbb{R}^4$, where each signal corresponds to a particular identity. In particular, each particle emits signals $\boldsymbol{\psi}$ as a function of their internal states $\mu$:

$$\boldsymbol{\psi} = \sigma(\mu) \qquad \text{(eq.1)}$$

Where $\sigma$ is the softmax function. The internal states $\mu \in \mathbb{R}^4$ of a particle encode model evidence (negative free-energy) for their identity, such that the distribution $\boldsymbol{\psi}_i$ represents the beliefs of a particle about its own identity. Particles thus transmit beliefs about their respective identity. These transmitted signals are sensed by neighbouring particles. Signals decay exponentially as a function of distance in their shared phase space (e.g., as in local vs. long-range synaptic connections within a cortical region or diffuse neuromodulator release), such that the local signal intensity $m \in \mathbb{R}^4$ for particle is given by:

$$m = \sum_{j}^{N} \boldsymbol{\psi}_j \exp\left(-\frac{1}{2}\,\Delta\vec{x}_j^2\right) \qquad \text{(eq. 2)}$$

where the vector between particle locations is given by $\Delta\vec{x}_j = \vec{x}_j - \vec{x}$.

We introduce two numerical innovations with respect to the simulations reported by Palacios and colleagues (2020). Rather than particles sensing local signal intensities directly, they sense the (2D) spatial flow vector of the four signals, $\vec{R} \in \mathbb{R}^2 \times \mathbb{R}^4$, the relative rate of change in the signal as a function of location, or its proportional intensity gradient:

$$\vec{R} = \frac{\nabla_x m}{m} \qquad \text{(eq. 3)}$$

By combining equations (1-3), we can show that the flow vector of any exponentially decaying variable satisfies:

$$\vec{R} = \sum_{j \neq i}^{N} \Delta\,\vec{x}_j \;\; \sigma\left(\mu_j - \frac{1}{2}\Delta\vec{x}_j^2\right) \qquad \text{(eq. 4)}$$

This equation shows that, for any particle, the flow vector $\vec{R}$ points directly to the (intensity-weighted) signal source. Moreover, the magnitude $R = |\vec{R}|$ of this vector is proportional to the distance from the signal source. These are remarkable properties for a signal that can always be computed locally, given the reasonable assumption of an exponential decrease of signal intensity with distance.

As well as sensing the flow vector magnitude $s_R$, each particle internally senses the signal it transmits $s_\psi$, such that the complete sensory state $s$ is:

$$s = \begin{pmatrix} s_\psi \\ s_R \end{pmatrix} = \begin{pmatrix} \boldsymbol{\psi} + \Omega_\psi \\ R + \Omega_R \end{pmatrix} \qquad \text{(eq. 5)}$$

where $\Omega_\psi$ and $\Omega_R$ are Gaussian fluctuations.

To enable self-organising behaviour, we equip the particles with prior beliefs about the magnitude of flow vectors $R$. As Bayesian agents act to realise prior beliefs (i.e., maximise model evidence), expectations about $R$ will promote behaviour (i.e., movement in state space) that realises the corresponding belief about distance from the intensity-weighted signal source.

We specify a unique prior $P(R|\psi)$ for each signal identity $\psi$, meaning that the signals (flow vector magnitudes) that are expected by each particle depend on their inferred identity:

$$P(R|\psi) = \mathcal{N}(\mathbf{R}_\psi, \Pi_R) \qquad \text{(eq. 6)}$$

where $\mathbf{R}_\psi \in \mathbb{R}^4$ is a set of four conditional means for each signal, given a particular identity $\psi$ and given the corresponding set of precisions $\Pi_R$ (inverse variances). Because $R$ is directly proportional to the (intensity-weighted) distance to the signal source, the Markov blanket structure of the superordinate ensemble can be encoded in these identity-specific expectations $\mathbf{R}_\psi$:

$$\mathbf{R}_\psi = \begin{bmatrix} \mathrm{D} & \mathrm{D} & 2\mathrm{D} & 2\mathrm{D} \\ \mathrm{D} & \mathrm{D} & \mathrm{D} & 2\mathrm{D} \\ 2\mathrm{D} & \mathrm{D} & \mathrm{D} & \mathrm{D} \\ 2\mathrm{D} & 2\mathrm{D} & \mathrm{D} & \mathrm{D} \end{bmatrix} \qquad \text{(eq. 7)}$$

Where $D$ is a scaling parameter. Effectively, this matrix encodes the compatibility of spatial proximity between the four identities. Specifically, particles can (infer they) belong to one of two higher-order ensembles, where these ensembles are themselves superordinate particles consisting of internal and blanket states. Particles can form beliefs about whether they are an internal state $\psi_1$ or a blanket state $\psi_2$ of one superordinate Markov blanket, or alternatively, a blanket state $\psi_3$ or an internal state $\psi_4$ of the alternative superordinate Markov blanket. These identities differ only in terms of their associated expectations $P(R|\psi)$, where inferring that one is an internal state $\psi_1$ of one Markov blanket involves the expectation of being far away (distance of 2D) from particles that are part of the other superordinate particle, $\psi_{3,4}$, and vice versa. Particles which identify as $\psi_2$ or $\psi_3$ correspond to the blanket states that separate the internal states of the superordinate ensembles. Accordingly, blanket states $\psi_2$ are compatible with proximity (distance of 1D) to the corresponding internal states $\psi_1$ as well as the blanket states $\psi_3$ of the other ensemble. Reciprocally, blanket states $\psi_3$ are compatible with proximity to internal states $\psi_4$ and blanket states of the other ensemble $\psi_2$.

The agents have a categorical belief distribution over identities and their actual expectations are a Bayesian model average of the expectations corresponding with each identity:

$$\mathbf{R} = \mathrm{E}_{\psi}[R] = \boldsymbol{\psi}\mathbf{R}_{\psi} \qquad \text{(eq. 8)}$$

We now derive updates for the model evidence for internal states $\mu$. The flow of the internal states is given by gradient flows on variational free-energy $F$:

$$f_{\mu}(\tilde{s}, \tilde{a}, \tilde{u}) = \left(Q_{\mu} - \Gamma_{\mu}\right) \nabla_{\mu} F \qquad \text{(eq. 9)}$$

Here, the tilde indicates generalised coordinates, containing both the variable and all its temporal derivatives. We refer readers to Palacios and colleagues (2020) for a full description. We consider a first-order approximation for small Gaussian fluctuations; in which case we obtain the rate of change of model evidence $\mu$:

$$\dot{\mu} = \nabla_{\mu}\mathbf{R} \cdot \Pi_R \varepsilon_R + \nabla_{\mu}\boldsymbol{\psi} \cdot \Pi_{\psi}\varepsilon_{\psi} - \Pi_{\mu}\mu \qquad \text{(eq. 10)}$$

Here, $\Pi_R$, $\Pi_{\psi}$, $\Pi_{\mu}$ are sensory and internal precisions (inverse variances) that normalise model evidence $\mu$ and the prediction errors $\varepsilon_R$, $\varepsilon_{\psi}$:

$$\begin{aligned} \varepsilon_{\psi} &= s_{\psi} - \boldsymbol{\psi} \\ \varepsilon_R &= s_R - \mathbf{R} \end{aligned} \qquad \text{(eq. 11)}$$

In other words, the model evidence for each particle identity varies with the normalised prediction error, where the prediction error quantifies the difference between the (magnitude of the) sensed flow vector and the prior beliefs about the magnitudes of flow vectors.

Finally, we move on to consider active states, which correspond to a chemical emitter ($a_{\psi} \in \mathbb{R}^4$ ) and motor actuators ($a_x \in \mathbb{R}^2$) that move the particle through phase space $x$. The flow of active states mirrors that of internal states:

$$f_{\tilde{a}}(\tilde{s}, \tilde{a}, \tilde{u}) = (Q_a - \Gamma_a) \nabla_{\tilde{a}} F \qquad \text{(eq. 12)}$$

Intracellular self-communication serves as a minimal type of memory that is regulated by a chemical emitter that communicates identity beliefs with normalised Gaussian fluctuations (mentioned above):

$$\begin{aligned} a_{\psi} &= \boldsymbol{\psi} \\ \dot{a}_{\psi} &= -\Pi_{\psi}\varepsilon_{\psi} \end{aligned} \qquad \text{(eq. 13)}$$

As discussed, motor actuators regulate the spatial location of cells. We still neglect friction for now ($\Pi_x = 0$), but we increase the realism of cell movement by choosing for a second-order approximation: $\tilde{a}_x = \binom{x}{\dot{x}}$, where $\dot{x}$ is the velocity. Since we are calculating the flow of both position and velocity as defined by the free-energy functional, this model also incorporates acceleration:

$$\begin{aligned}\dot{x} &= -\nabla_x \vec{R} \cdot \Pi_R \varepsilon_R \\ \ddot{x} &= \nabla_x \vec{R} \cdot \lambda \Pi_R \dot{\mathbf{R}}\end{aligned} \qquad \text{(eq. 14)}$$

where λ parametrises the autocorrelation of fluctuations over time, which sets the smoothness of fluctuations and regulates the magnitude of second-order effects (see Friston et al., 2010 for its derivation) and $\dot{\mathbf{R}}$ is the temporal derivative of expectations about $R$:

$$\dot{\mathbf{R}} = \mathbf{R}\boldsymbol{\psi}(1-\boldsymbol{\psi})\dot{\mu} \qquad \text{(eq. 15)}$$

Collecting all results, we can now integrate over time and use a Taylor series expansion to simulate trajectories of particles over time in terms of $x$, $a_\psi$ and $\mu$:

$$\begin{aligned}x_{t+1} &= x_t + \dot{x}_t \Delta t + \frac{1}{2}\ddot{x}_t \Delta \mathrm{t}^2 \\ a_{\psi,t+1} &= \boldsymbol{\psi} + \Pi_\psi \Omega_\psi \Delta t \\ \mu_{t+1} &= \mu_t + \dot{\mu}_t \Delta t\end{aligned} \qquad \text{(eq. 16)}$$

In summary, all we have done here is write down a relatively simple generative model of what any neuron or particle expects to sense, depending upon its identity (e.g., functional selectivity) and spatial relationship to other particles that share the same generative model. This generative model means that one can compute variational free-energy gradients and the ensuing flow of particular states, which can be read as neuronal dynamics or movement. This dynamics will necessarily show self-organising, self-assembling, and self-evidencing aspects because it minimises a variational bound on the evidence for a (shared) model of the external milieu. Crucially, the external milieu is made up of particles that are all trying to do the same thing and thereby fulfil each other's expectations. Technically, because we are dealing with continuous states and Gaussian fluctuations, the variational free-energy gradients become prediction errors. This can be seen in the above equations where the flow of internal and active states is a function of various prediction errors. Biophysically, these prediction errors would correspond to the setpoints of electrochemical variables, such that the particles keep moving when prediction errors are all zero and each particle has found its variational free-energy minimum.

In order to illustrate the concepts outlined in the previous sections, we constructed two sets of numerical experiments. These will illustrate, respectively:

- The emergence of a two Markov blanketed systems, which results from the application of an external field that provides evidence of ensemble membership ($\psi_{1,2}$) to a small subset of particles.
- The transient and dynamic nature of Markov blanket formation in response to changing external fields.

## 4. Results

### 4.1. The emergence of a neuronal packet

The results of the first series of numerical simulations are presented in Figure 4. In these simulations, a subpopulation of the individual neurons was stimulated to induce the precise posterior belief that they were states of one Markov blanketed system at the superordinate level ($\psi_{1,2}$). This was achieved by coupling the selected neurons to an external field, which, as described as above, is a metaphor for signals received from neurons in the sensory epithelia of the organism.

We introduce a stimulus function $f_{stim}(x) \to \mathbb{R}^1$ that specifies the stimulus intensity as a function of phase space position. This function determines the evidence a neuron receives for being a member of the first ensemble $\psi_{1,2,}$ while the inverse of this function $1 - f_{stim}(x)$ determines the evidence a neuron receives for the alternative ensemble $\psi_{3,4}$.

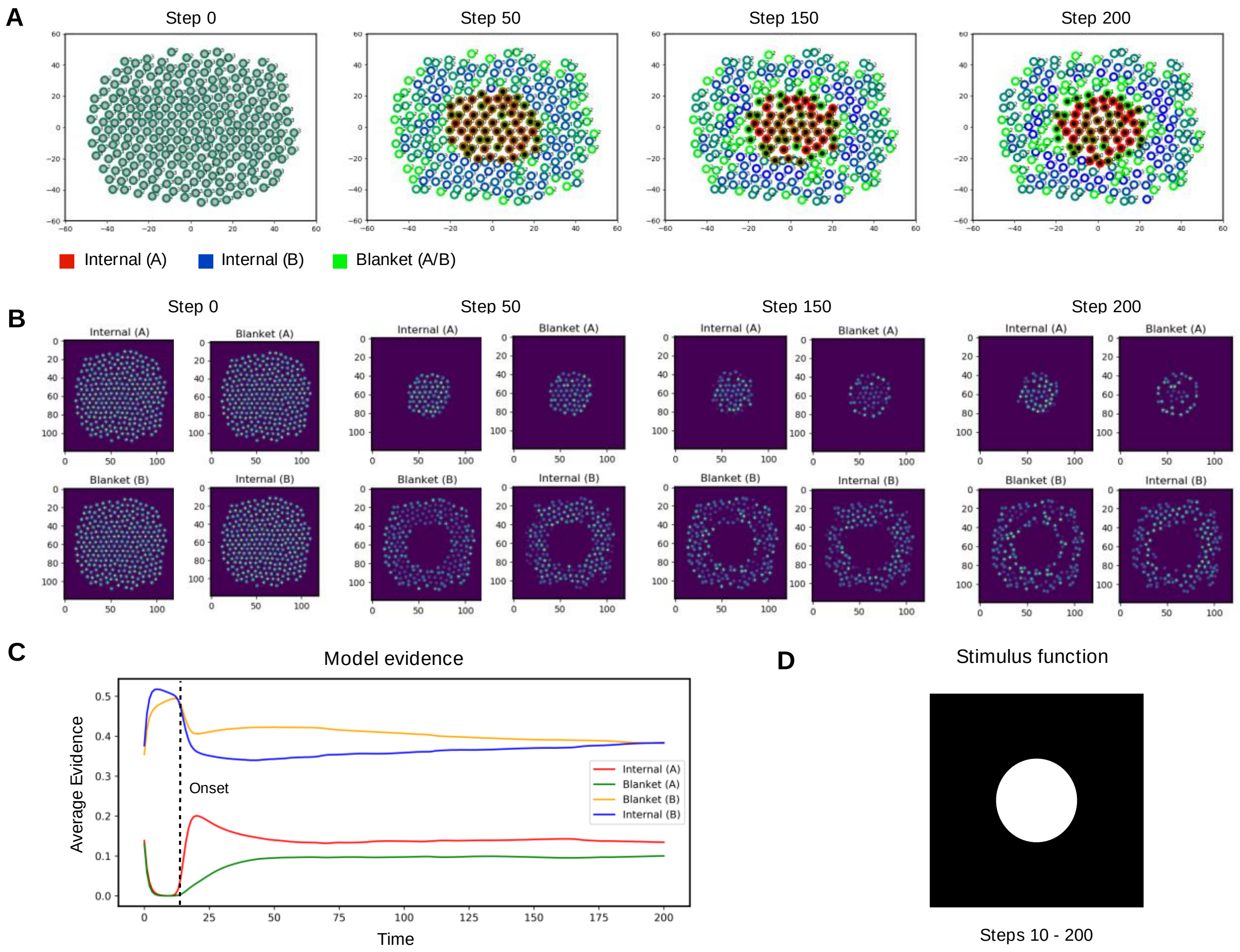

**Figure 4.** The emergence of a neuronal packet.
This figure depicts the first series of numerical results, which demonstrate the emergence of a Markov blanket in response to external stimulation – or coupling with an external field. **(A)** A visualisation of individual neurons at different times during the stimulation. The colour of a neuron represents its beliefs about membership in superordinate Markov blankets, while its location denotes its position in phase space. Black and white represent distinct Markov blankets; while line colour represents role in the Markovian partition (as internal or blanket states of the respective blanketed ensemble). At the start of the stimulation, neurons are initialised with uniform beliefs about their identity. At time step $T = 10$, an external stimulus is introduced (described in the main text). The external stimulation causes neurons to divide into two sets of internal and blanket states, organised around the stimulus. Over time, shared expectations about what each neuron should sense cause the neurons to divide into internal and blanket states. The internal states of the first ensemble (A) organise around the stimulus centre, while the blanket states of this ensemble organise around the edges of the stimulus. Of the remaining neurons, those which are in spatial proximity to the first ensemble (A), or are at the external boundary of the ensemble, form beliefs that they are blanket states of the second ensemble (B), while the other neurons form beliefs that they are internal states of that ensemble. The external stimulation thus induces the formation of two Markov blankets, the internal states of which are conditionally independent of one another, conditioned on the blanket states. **(B)** A visualisation of the signals emitted by neurons, averaged over the ensemble, for each location in phase space. In panels (A) and (B), the two axes represent the two abstract states of a two-dimensional state space. For simplicity, here, we assume this is a physical space where the agents move, with real values $\mathbb{R}^2$. **(C)** The ensemble-average evidence for each type of identity as a function of time. **D)** The stimulus function used in this simulation, where the white denotes the stimulus intensity.

At the start of the simulation, neurons were initialised with uniform beliefs about their identity. The stimulus field was then introduced at time step $T = 10$, causing the neurons within this field to quickly form beliefs that were members of the first superordinate ensemble ($\psi_{1,2}$). Conversely, the remaining neurons formed beliefs that they are members of the second superordinate ensemble ($\psi_{3,4}$). Over the course of the simulation, the shared expectations about proximity relations $\mathbf{R}_\psi$ caused neurons to converge to either internal ($\psi_{1,4}$) or blanket ($\psi_{2,3}$) states. Specifically, neurons which were not in spatial proximity to states of the opposing higher-level ensemble formed beliefs that they were internal states, whereas neurons which were in spatial proximity to both ensembles formed beliefs that they were blanket states. This caused the first Markov blanket to form around the field introduced by the stimulus function, and the second Markov blanket to envelope the first. The emergence of this structure meant that when conditioned on their blanket states $\psi_2$, the internal states of the first Markov blanket $\psi_1$ were independent of the states of the second Markov blanket $\psi_{3,4}$, and vice versa.

These results may also speak to the role of local field potentials in neural synchronization. The prior belief that they will synchronise with each other in the absence of stimulation lends to the individual neurons that make up a neuronal system a kind of coherence, which gives it the look and feel of a fictitious ‘centripetal’ force. Conversely, when neurons are stimulated by contradictory driving inputs, they will self-organize into distinct and segregated groups, which will look like a ‘centrifugal’ force. The group inference that realizes population coding effectively balances these two opposing tendencies in self-organization.

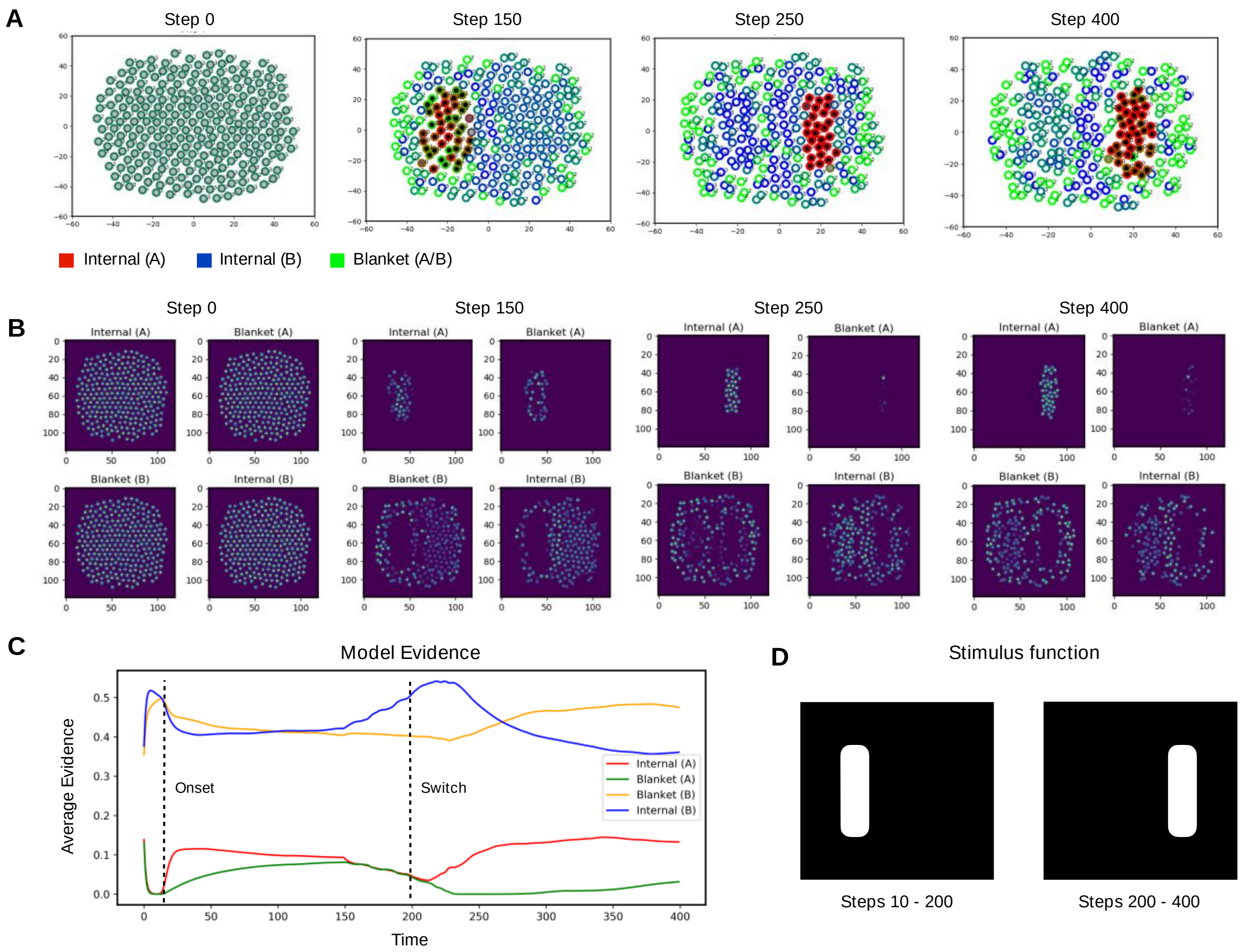

**Figure 5.** Dynamic and transient packet formulation
This figure depicts the second series of numerical results, which demonstrate the dynamic nature of Markov blanket formulation in response to transient external stimulation. **(A)** A visualisation of individual neurons at different times during the stimulation. As in the first set of simulations, the neurons are initialised with uniform beliefs and at time step $T = 10$ an external stimulus is introduced (described in the main text). Following the results of the first simulation, the external field induces the formation of two Markov blanketed systems organised around the stimulus. At time step $T = 200$, the location at which the stimulus is applied is changed. This leads a subpopulation of neurons to receive evidence that is incompatible with their Bayesian beliefs, causing the dissolution of the original Markov blanket and the emergence of a new Markov blanket around the changed site of stimulation. These results demonstrate the dynamic and transient nature of Markov blanket formation when coupled to an external field. **(B)** A visualisation of the signals emitted by neurons at different times during the stimulation. Again, in panels (A) and (B), the two axes represent the two abstract states of a two-dimensional state space. For simplicity, here, we assume this is a physical space where the agents move, with real values $\mathbb{R}^2$. **(C)** The ensemble-average evidence for each type identity as a function of time. **D)** The stimulus function used in this simulation. As above, white denotes the stimulus intensity. Here, the stimulus changes location at 200 time steps.

## 4.2. Splitting ensembles and multiple ensembles: Towards an account of dynamic, transient Markov boundary formation

We move on to consider dynamic nature of Markov blanket formation in response to changing external stimulation. Following the first set of simulations, we initialise neurons with uniform beliefs about their identity and introduce an external stimulus at $T = 10$. However, we now transform the external stimulus at $T = 200$. This change in stimulation causes a subpopulation of neurons to form incompatible beliefs and promotes the reorganisation of the superordinate ensembles.

The results of these simulations are shown in Figure 5 and 6, which differ only in the stimulus function used. As with the first set of simulations, these results show the emergence of two Markov blanketed systems around the initial stimulus. However, the results additionally demonstrate that once the location of the stimulus changes, the entire ensemble self-organizes anew. Specifically, a new Markov blanket organises around the location of the new stimulus, which is a direct result of the evidence provided by the stimulus field. The new blanket forms on a slower time scale to that of the original, due to the fact that members of this new ensemble originally believed they were members of the alternative ensemble ($\psi_{3,4}$). Moreover, since the original external field has been removed, the corresponding neurons no longer receive evidence that they are states of the first ensemble ($\psi_{1,2}$). Instead, they receive evidence that they are members of the opposing ensemble ($\psi_{3,4}$), leading to the gradual dissolution of the first Markov blanket. Taken together, these results demonstrate the dynamic nature of Markov blanket formation when coupled to a dynamic external field.

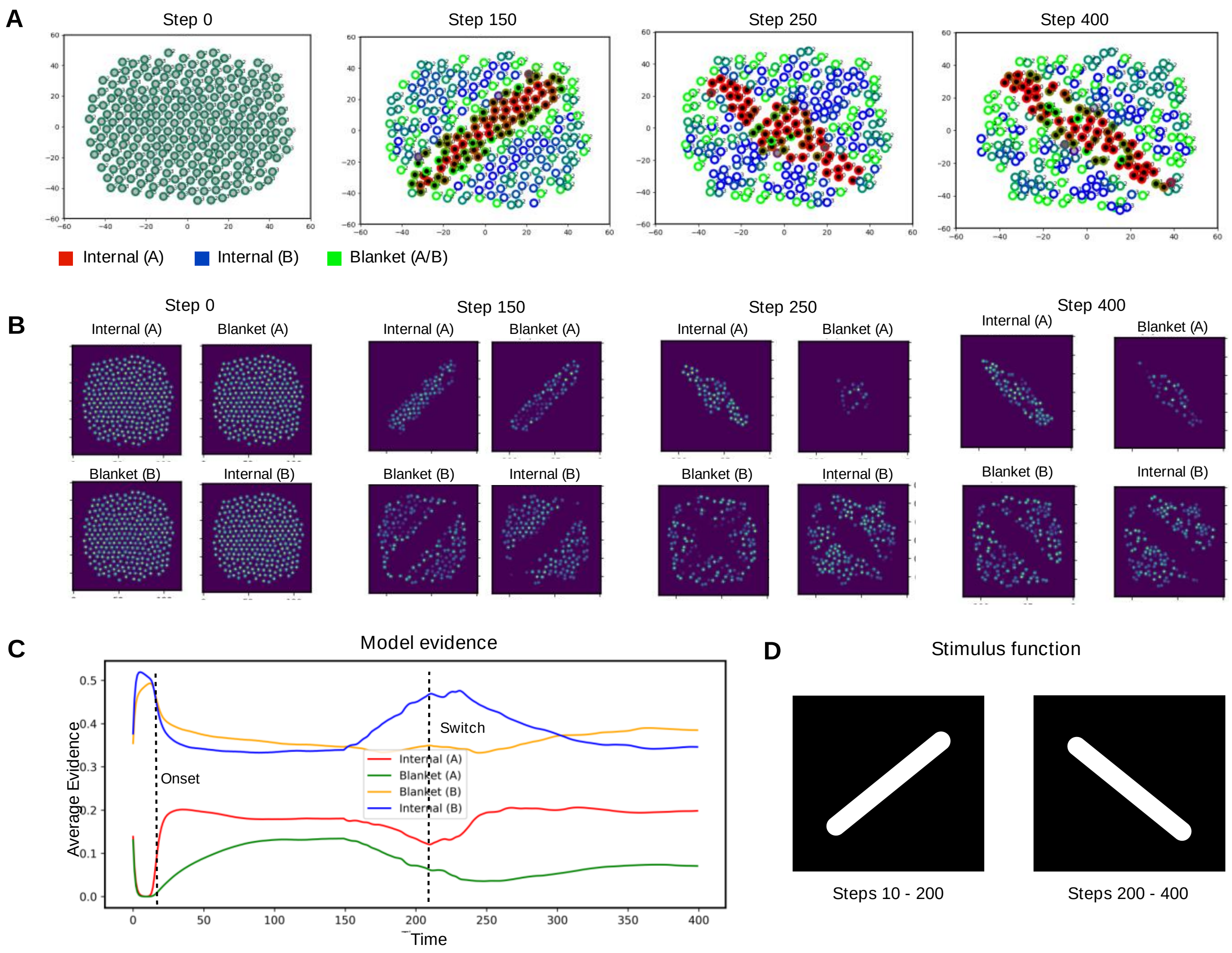

**Figure 6.** Dynamic and transient packet formulation (B)
This figure depicts the final series of numerical results, which further demonstrates the dynamic nature of Markov blanket formation in response to external stimuli. These numerical results utilise a different set of set stimuli relative to Figure 5. **(A)** A visualisation of individual neurons at different times during the stimulation. As in the previous simulations, neurons are initialised with uniform beliefs and at time step $T = 10$ an external stimulus is introduced. This again induces the formation of two Markov blanketed systems organised around the induced stimulus. At time step $T = 200$, the location at which the stimulus is applied is changed, and this again causes the location of Markov blankets to dynamically reorganise. Unlike in Figure 5, the two stimuli overlap, meaning a subset of the population receives consistent evidence after the stimulus is changed. These beliefs of these neurons do not change, leading to faster blanket formation, relative to Figure 5. **(B)** A visualisation of the signals emitted by neurons at different times during the stimulation. As above, in panels (A) and (B), the two axes represent the two abstract states of a two-dimensional state space. For simplicity, here, we assume this is a physical space where the agents move, with real values $\mathbb{R}^2$. **(C)** The ensemble-average evidence for each type identity as a function of time. **D)** The stimulus function used in this simulation. As above, white denotes the stimulus intensity. Here, the stimulus changes location at 200 time steps.

## 5. Discussion

This paper has presented a model of neural and phenotypic representation developed using the active inference framework. One potential advantage that our model offers is that it builds on the current neural process theory associated with active inference. The current process theory models single-neuron activity – which can also simply be equated with activity in populations of neurons – as more or less uniformly encoding posterior beliefs over states; moreover, it assumes that the synaptic connection patterns giving rise to the selective stimulus sensitivity of a given neuron (or population) are formally specified as single entries in a likelihood matrix connecting sensory epithelia to those neurons. Heuristically, learning in these models is Hebbian, in that synaptic connections in the likelihood matrix are strengthened based on the strength of the activity of a neuron (or population), coincident with current observations.

The key points are that – in extant models – no distinction is made between the function of a single neuron versus a population of neurons encoding posterior beliefs over states, or learning stimulus sensitivity, and that the activation levels that are modelled are single trajectories of values. Our results in this paper demonstrate that coherent activity in neural populations can arise in response to environmental perturbation and allow for population-wide tuning of synaptic connections during development – such that different populations of neurons could learn differential sensitivities to different stimulus features. This avoids possible concerns about modelling 'grandmother cells' – and may offer a means of redundant coding that allows a system to be robust to noise or injury.

On a related note, the formalism used to model the emergence of a selective responses and specialisation (e.g., receptive fields) – as a free energy minimising process – may resolve arguments about whether probabilistic representations are represented by the sample density of population responses; i.e., a population code in the sense of Beck et al. (2008); Pouget, Beck, Ma, and Latham (2013); Zemel, Dayan, and Pouget (1998) – or whether the population response encodes the parameters or sufficient statistics of a probabilistic representation; i.e., an ensemble code. On the present account, the two are the same thing. In other words, the quantity that matters – in terms of a representational or extrinsic information geometry – is the ensemble average of a population of neurons that constitute the internal states of a larger Markov blanket.

A subtle, technical point here is that the extrinsic information geometry that undergirds neuronal representation or encoding rests upon the conditional expectation or average of internal states, e.g., neurons that are internal to a larger Markov blanket. However, mathematically, this expectation is over time, not over internal states (Friston, 2020). To endow ensemble averages with a representational attribute, it then becomes necessary to make the ergodic assumption that the average over a large number of internal states is the same as the time average. This only works if the internal states are exchangeable or can be treated as equivalent to each other. This simple observation means that there may be something quintessential about population codes; in the sense that they can only be attributes of ensembles, e.g., populations of neurons that play the role of internal states at a superordinate scale.

This work presented here also demonstrates novel Markov blanket dynamics, opening the door to the modelling of transient boundary formation and reconfiguration – and dynamics in functional and effective connectivity – consistent with a growing body of evidence supporting neural reuse (Anderson, 2014). Some systems have permanent boundaries that endure over time. This is the case, for instance, when the statistical boundaries of a system coincide with its physical ones, as in the case of a cell, where the statistical boundary is also hardwired into the physical structure of the system, as a literal barrier (a cell membrane). Other living systems, however, have more malleable boundaries, which dissolve and reform over time. These are systems like social systems (Veissière, Constant, Ramstead, Friston, & Kirmayer, 2020) and – arguably – neuronal packets (Yufik & Friston, 2016). When we wake up in our home, we are part of the intimate living unit; when we go to the office, we are part of another functional unit; and belonging to these groups constrains possible courses of action. This is also the case for neural ensembles, which seem to form and reform on the fly to match task requirements (Anderson, 2014). For systems such as these, the relevant boundaries are functional precisely because they can change, as when one individual member moves to another group, when a group splits into two, or when two groups merge. The boundaries of such neural (or social) systems are intrinsically malleable, which allows for adaptive reconfigurations of groups and group membership. We propose that one can associate the changing architecture and membership of superordinate ensembles with inference dynamics that track different perceptual contents.

Note that fluctuating and itinerant Markov blankets are consistent with non-equilibrium steady-state, due to a separation of temporal scales that emerges with Markov blankets of Markov blankets (Friston, 2020). In other words, any Markov blanket will change slowly from the point of view of Markov blankets at the scale below. For example, the Markov blanket of a city does not change hour by hour, as you commute from work to home. Similarly, the Markov blanket of your brain does not change second by second, as neuronal ensembles (i.e., packets or assemblies) split and merge. The key thing – from the perspective of ensemble dynamics – is that various (attracting) states are revisited after a sufficient period of time. For example, you return home every evening. And your functionally specialised neuronal ensembles are activated every few seconds by characteristic patterns of sensory afferents.

The framework that we develop here has implications that transcend neuronal dynamics. Indeed, the scheme offers a formal, generalizable account of phenomena that involve the formation of transient boundaries in nested systems. The ensuing formalism provides a generic and general mechanism by which coordinated movement in living creatures might act as an effective means to model or represent the causal situations, environments, and predicaments in which living creatures find themselves.

Group foraging behaviours, for instance, obey similar principles, where the coordinated patterns of movement in a direction count as accumulated evidence for specific hypotheses about the structure of the environment. This is the case, for instance, when ants leave and follow pheromone trails. In this setup, the more ants that follow a given path, the more evidence each provides to its conspecifics for the belief that there is something salient (a food item or danger) at the end of the trail (Jackson, Holcombe, & Ratnieks, 2004).

We suggest that it may be possible to extend the modelling strategy developed above to account for the ways in which social and cultural group membership functions as a form of inference about relevant causal structure in the human environment. For instance, political affiliation seems to track one's beliefs and attitudes regarding some domains, such as the reality of human-made climate change. One might speculate that for many human agents, the individual inference problem (namely, to answer for oneself the question, 'Do I believe that human-made climate change is real?') is solved by outsourcing the process of generating a working guess (computing the posterior) to one's close social circle, often by figuring out and learning those things believed and valued in one's social ingroup (Kirmayer & Ramstead, 2017; Ramstead, Veissière, & Kirmayer, 2016; Veissière et al., 2020).

Another, less politically contentious example is the optimal foraging behaviour in ant colonies discussed above. Indeed, ant foraging behaviour might be modelled as a kind of group inference, where each foraging 'arm' of the ant population (enshrouded behind its own Markov blanket, here determined by chemical pheromone gradients ) can be cast as representing a distinct hypothesis that 'there is food here', or that 'there is a dangerous threat here', and so on. One advantage of applying this formalism to such animal models -- instead of human systems -- is that they are comparatively easier to study, since they are simpler to extensively (and ethically) measure and since their (relatively less complex) behaviour is easier to model. Work on ant foraging and on other simple foraging systems, such as the multiple search fronts of foraging slime mould populations, routinely employs modelling strategies amenable to active inference (Chandrasekhar, Gordon, & Navlakha, 2018; Reid, Latty, Dussutour, & Beekman, 2012). Behavioural data regarding the decisions made by those simple systems is equally reproducible in silico.

Finally, one might be able to extend the above population coding strategy to *hierarchically nested* representational capacities. In such a scheme, neuronal packets at the superordinate level would engage in exactly the same kind of coordinated inferential dynamics, inferring their membership identity in yet a higher level of self-organization. This could lead from a theory of neural representation to a theory of conceptual- or semantic-level processing (Yufik, 2019; Yufik & Friston, 2016) – as opposed to the perceptual processing formally modelled here – where neuronal packet formation at several nested scales in the brain explains the perception and multimodal integration of the perception of objects. Indeed, current formulations of understanding and abstraction rest upon deep generative models with relational structure that necessarily call upon Markov blankets to separate hierarchical levels; e.g., Friston, Lin, et al. (2017). The time and context sensitive nature of Markov blankets – considered in this work – may reflect the context sensitive, state dependent, nonlinear generative models that are necessary for self-evidencing.

## 6. Conclusion

The aim of this paper was to develop a generic and generalizable model of the representational capacities of living creatures by using the resources of the free-energy principle and its corollary process theory, active inference. We developed an account of neuronal and, more broadly, phenotypic representation. We argued that the representational capacities of living creatures are a consequence of their Markovian structure and action in nonequilibrium regimes to counter the dissipating effects of random fluctuations and environmental itinerancy. We demonstrated *in silico* that information about external stimuli can be encoded by groups of neurons bound by a Markov blanket, thereby providing evidence for the *neuronal packet hypothesis* that inherits from theories of neuronal groups and assemblies (Buzsaki, 2010; Edelman, 1993; Edelman, 1998; Hebb, 1949; von der Malsburg, 1981). Our numerical proof-of-principle simulations showed that self-organizing ensembles of particles that share the right kind of probabilistic generative model are able to encode recoverable information about a changing sensorium – and that they can dynamically evince patterns of functional self-organization that adapt to the demands of the situation. This model of neuronal packet formation thus entails a modest form of representational capacity and opens up important directions for future work.

## Software note

The Python code for this paper is available at https://github.com/alec-tschantz/packets. To recreate the figures four, five and six, please run packets_figure_01.py, packets_figure_02.py and packets_figure_03.py, respectively. Further information on recreating the figures and data are available in the linked repository. This code is based on a generalised filtering scheme written in the SPM software for the MATLAB environment; namely, the DEM_cells.m and DEM_cells_cells.m scripts, which implement the self-organisation of a single ensemble Markov blanketed ensemble and an ensemble of such ensembles, respectively. The original routines are available at http://www.fil.ion.ucl.ac.uk/spm.